\documentclass[aps,prl,twocolumn]{revtex4}

\usepackage{graphicx}
\usepackage{amsmath}

\begin{document}

\title{Effective capacitance of a single-electron transistor}

\author{M.~A. Laakso}
\author{T. Ojanen}
\author{T.~T. Heikkil\"a}
\email[]{Tero.Heikkila@tkk.fi}
\affiliation{Low Temperature
Laboratory, Helsinki University of Technology, P.O. Box 5100
FIN-02015 TKK, Finland}

\date{\today}

\begin{abstract}
Starting from the Kubo formula for conductance, we calculate the frequency-dependent response of a single-electron transistor (SET) driven by an ac signal. Treating tunneling processes within the lowest order approximation, valid for a wide range of parameters, we discover a finite reactive part even under Coulomb blockade due to virtual processes. At low frequencies this can be described by an effective capacitance. This effect can be probed with microwave reflection measurements in radio-frequency (rf) SET provided that the capacitance of the surroundings does not completely mask that of the SET.
\end{abstract}

\pacs{72.10.-d,73.23.Hk,72.30.+q}

\maketitle

\begin{figure}
 \centering
 \includegraphics[width=8cm]{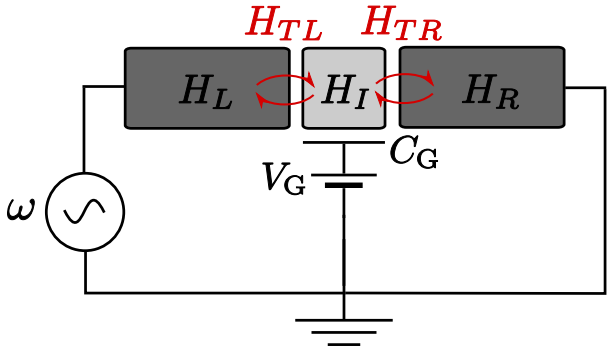}
 \caption{(color online) Schematic drawing of a single-electron transistor considered in this Letter. Left electrode is connected to an ac voltage source of angular frequency $\omega$. Central island is coupled capacitively to a gate electrode, and the gate charge can be adjusted with the gate voltage $V_G$.}
 \label{fig:schema2}
\end{figure}

A single-electron transistor (SET), shown schematically in Fig.~\ref{fig:schema2}, is one of the most widely studied components of nanoelectronics today. Numerous applications include charge detection in mesoscopic structures \cite{rfset}, thermometry \cite{cbt} and single-electron pumping \cite{pothier}. Due to its high sensitivity to charging effects, a SET is an ideal structure to study and characterize single-electron and quantum effects such as Coulomb blockade and tunneling.

The orthodox theory developed by Averin and Likharev \cite{averin+likharev} has been extremely successful in describing the dc properties of a SET. More recently, however, SET has also been used at finite frequencies as an accurate electrometer in the form of rf-SET \cite{rfset}. In this application, the impedance of the SET has been assumed to be completely resistive. Previously the response of a SET to an ac field has been studied in the framework of scattering matrix formalism \cite{buttiker+pretre,buttiker+pedersen} and with the Tien-Gordon approach \cite{aguado,flensberg,kouwenhoven}. In this Letter we utilize the Kubo linear response formula to calculate the finite-frequency admittance of a SET in the presence of Coulomb interaction. In addition to the familiar sequential tunneling effects, at finite frequencies the lowest order approximation describes some virtual tunneling processes as well. These give rise to a nonzero reactive response even under Coulomb blockade. At low frequencies this contribution is described by an effective capacitance, which can be tuned with the gate voltage. We also discuss the possibility of detecting this gate-dependent capacitance through the tuning of the resonant frequency in an LC circuit.

Usually it is the familiar geometric capacitance that dominates the total capacitance of a given system. There are, however, additional contributions that have a microscopic origin. These are the density of states capacitance \cite{datta, sillanpaa, ilani}, which results from the finite amount of kinetic energy that is required to introduce an additional electron to a conductor, and the correlation capacitance, which comes from the correlated motion of electrons.

The Hamiltonian for the single-electron transistor is $H=H_L+H_I+H_R+H_T$, where $H_L=\sum_{\nu}\varepsilon_{\nu}c^\dagger_{\nu}c_{\nu}$ describes the noninteracting electrons with eigenstates $\nu$ in the left lead and $H_R$ is a similar expression for electrons in the right lead. The island is described with the Hamiltonian
\begin{equation}
 H_I=\sum_{\mu}\varepsilon_{\mu}c^\dagger_{\mu}c^{\phantom{\dagger}}_{\mu}+E(N), \quad
 E(N)=E_CN^2-eV_GN,
\end{equation}
where $N$ is the number of excess electrons on the island and $E_C=e^2/2C$ is the charging energy for an island of capacitance $C=C_L+C_R+C_G$. For the sequel, it is useful to define $\delta E^\pm_N=E(N\pm1)-E(N)=(\pm2N+1)E_C\mp eV_G$. The tunneling Hamiltonian, describing the charge transfer processes between the island and the leads, is of the usual form
\begin{align}
H_T=&H^+_{TL}+H^-_{TL}+H^+_{TR}+H^-_{TR}, \nonumber\\
H^+_{T\alpha}=&\sum_{\nu\mu}t_{\nu\mu}c^\dagger_{\nu}c^{\phantom{\dagger}}_{\mu},
\quad H^-_{T\alpha}=\left(H^+_{T\alpha}\right)^\dagger,
\end{align}
where $t_{\nu\mu}$ is the tunneling matrix element between two corresponding states and $\alpha\in{L,R}$. The Kubo formula for conductance reads \cite{bruus+flensberg}
\begin{equation}\label{eq:kubo}
 G(\omega)=\frac{ie^2}{\omega}C^R_{II}(\omega)+\frac{ic_0}{\omega},
\end{equation}
where the retarded current-current correlation function is the Fourier transform of
\begin{equation}
 C^R_{II}(t-t')=-i\theta(t-t')\langle [I(t),I(t')]\rangle
\end{equation}
and the second term involving positive constant $c_0$ is the so-called diamagnetic term, which cancels the zero frequency divergence in the imaginary part of conductance \cite{haug+jauho}.

The particle current operator for a single-electron transistor can be found as the time derivative of the particle number in, say, the left lead. The Heisenberg equation of motion for the number operator yields
\begin{equation}
 \dot{N}_L(t)=i\left[H,N_L(t)\right]=i\left[H_T,N_L(t)\right],
\end{equation}
which leads to
\begin{equation}
 I(t)=\dot{N}_L(t)=-i\left(H^+_{TL}(t)-H^-_{TL}(t)\right).
\end{equation}
The current-current correlator is then
\begin{equation}
 C^R_{II}(t-t')=2\:\mathrm{Im}\:\theta(t-t')\left\langle\left[H^-_{TL}(t),H^+_{TL}(t')\right]\right\rangle,
\end{equation}
where the time-evolution of the operators is determined by the full Hamiltonian and the expectation value should be calculated in the presence of tunneling. Transforming this expression to the interaction picture and expanding to the lowest non-vanishing order in $t_{\nu\mu}$ we find
\begin{align} \label{eq:crii}
C^R_{II}(t-t')=&2\:\mathrm{Im}\:\theta(t-t')\sum_{\nu\mu}\sum_{\nu'\mu'}t^*_{\nu\mu}t^{\phantom{*}}_{\nu'\mu'}
\nonumber \\
&\times\langle[c^\dagger_{\mu}(t)c^{\phantom{\dagger}}_{\nu}(t),c^\dagger_{\nu'}(t')c^{\phantom{\dagger}}_{\mu'}(t')]\rangle_0,
\end{align}
where the subscript zero means that the expectation value and time-evolution of the operators should be evaluated with respect to the Hamiltonian $H_L+H_I+H_R$. Using standard methods of many-body theory, the expectation value in Eq.~\eqref{eq:crii} can be decomposed to Fermi functions and time-dependent exponentials with the help of the finite-temperature Wick's Theorem. Thus, in frequency domain Eq.~\eqref{eq:crii} takes the form
\begin{align} \label{eq:crii2}
&C^R_{II}(\omega)=\sum_{\nu\mu}\left|t_{\nu\mu}\right|^2\biggl[(1-f(\varepsilon_{\nu}))f(\varepsilon_{\mu})\biggr.
\nonumber \\
&\times\left(\frac{1}{\omega-\varepsilon_{\nu}+\varepsilon_{\mu}-\delta
E^-_N+i\eta}-\frac{1}{\omega+\varepsilon_{\nu}-\varepsilon_{\mu}+\delta
E^-_N+i\eta}\right) \nonumber \\
&-f(\varepsilon_{\nu})(1-f(\varepsilon_{\mu})) \nonumber \\
&\times\left.\left(\frac{1}{\omega-\varepsilon_{\nu}+\varepsilon_{\mu}+\delta
E^+_N+i\eta}-\frac{1}{\omega+\varepsilon_{\nu}-\varepsilon_{\mu}-\delta
E^+_N+i\eta}\right)\right],
\end{align}
where $\eta$ is a positive infinitesimal quantity. We transform the sum to an integral and using $1/(x+i\eta)=\mathcal{P}\:1/x-i\pi\delta(x)$ the imaginary part takes the form
\begin{align}\label{eq:imc}
\mathrm{Im}\:C^R_{II}(\omega)=&-\frac{1}{2e^2R_T}\left(\frac{\delta
E^-_N-\omega}{e^{\beta(\delta E^-_N-\omega)}-1}-\frac{\delta
E^-_N+\omega}{e^{\beta(\delta E^-_N+\omega)}-1}\right. \nonumber \\
&+\left.\frac{\delta E^+_N-\omega}{e^{\beta(\delta
E^+_N-\omega)}-1}-\frac{\delta E^+_N+\omega}{e^{\beta(\delta
E^+_N+\omega)}-1}\right),
\end{align}
and the real part, without the diverging term cancelled by the diamagnetic term in Eq.~\eqref{eq:kubo}, is given at $T=0$ by
\begin{align}\label{eq:rec}
&\mathrm{Re}\:C^R_{II}(\omega)=\frac{1}{2\pi
e^2R_T}\left\{\omega\ln\left|\frac{(\delta
E^-_N-\omega)(\delta E^+_N-\omega)}{(\delta E^-_N+\omega)(\delta
E^+_N+\omega)}\right|\right. \nonumber \\ &\left.+\delta
E^-_N\ln\left|\frac{(\delta E^-_N)^2}{(\delta E^-_N)^2-\omega^2}\right|+\delta
E^+_N\ln\left|\frac{(\delta E^+_N)^2}{(\delta E^+_N)^2-\omega^2}\right|\right\}.
\end{align}
Here $R_T=\hbar(2\pi e^2|t_{\nu\mu}|d_Ld_R)^{-1}$ is the usual tunneling resistance determined by the tunneling amplitude and densities of states $d_L, d_R$ at the Fermi level. Numerical integration of Eq.~\eqref{eq:crii2} shows that the temperature dependence of the real part is exponentially weak when $k_BT\ll E_C$. The conductance follows from Eq.~\eqref{eq:kubo}. Note that the real part of the conductance is obtained from the imaginary part of the current-current correlator and vice versa.

\begin{figure}
 \centering
 \includegraphics[width=8.5cm]{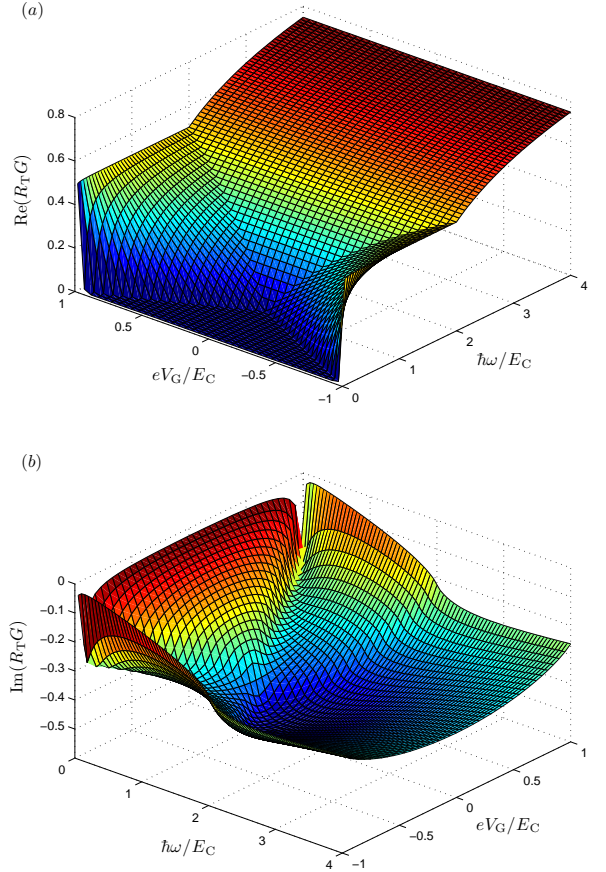}
 \caption{(color online) Real (a) and imaginary (b) part of the zero-temperature admittance as a function of the frequency of the driving signal and the gate voltage. If the SET is assumed to stay in the ground state at all times, the patterns repeat periodically as functions of the gate voltage.}
 \label{fig:conductance}
\end{figure}
Real part of the admittance, shown in Fig.~\ref{fig:conductance}(a), exhibits the familiar Coulomb blockade as can also be seen from Eq.~\eqref{eq:imc}. If the frequency of the driving signal is not sufficiently high to provide the required charging energy, real part of the admittance is exponentially suppressed and vanishes completely at zero temperature. For higher frequencies the blockade is lifted, which corresponds to photon assisted tunneling. When $eV_G=(2N\pm1)E_C$, two lowest charge states are degenerate, and the charging energy vanishes. Imaginary part of the admittance is shown in Fig.~\ref{fig:conductance}(b). It is nonzero even under Coulomb blockade because of the possibility of electrons to tunnel back and forth to a virtual state on the island. It should be noted that for dc-response it is necessary to take the second order approximation in $H_T$ for virtual processes to appear \cite{averin+nazarov}. The magnitude of the imaginary part grows linearly at low frequencies, as can be seen by expanding Eq.~\eqref{eq:rec} near $\omega=0$. We obtain a linear admittance
\begin{equation}
 \mathrm{Im}\:G\approx-\frac{\hbar\omega}{2\pi R_T}\left(\frac{1}{\delta E^-_N}+\frac{1}{\delta E^+_N}\right),
\end{equation}
which implies capacitive behavior with an effective capacitance
\begin{align}\label{eq:capacitance}
 \tilde{C}=&\left(\frac{e}{2\pi}\right)^2\left(\frac{R_Q}{R_T}\right) \nonumber \\ &\times\left(\frac{2E_C}{[(2N+1)E_C-eV_G][(1-2N)E_C+eV_G]}\right),
\end{align}
where $R_Q=h/e^2$ is the quantum resistance. Capacitance can be tuned with the gate voltage and it diverges at the charge degeneracy point. This divergence, appearing only at zero frequency, is an artifact of the first order approximation and is cancelled by higher order contributions. Similar result was found for a single tunnel junction under dynamical Coulomb blockade in Ref.~\cite{sonin}.

\begin{figure}
 \centering
 \includegraphics[width=8.5cm]{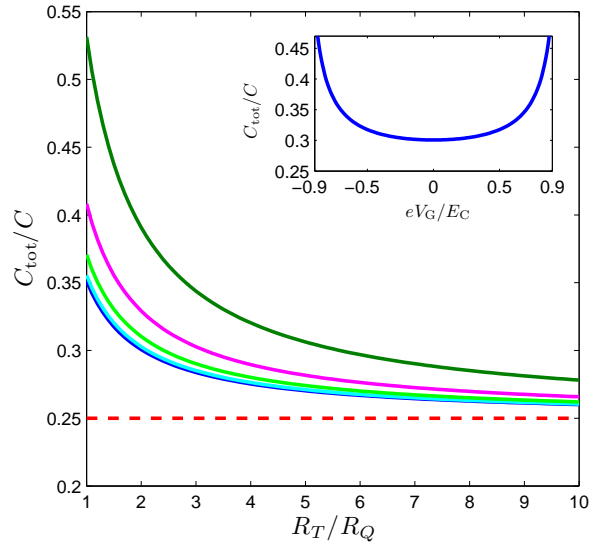}
 \caption{(color online) The total capacitance of the SET as a function of $R_T/R_Q$. The different curves correspond to different values of gate voltage (from top to bottom): $eV_G/E_C=0$ (blue), $eV_G/E_C=0.2$ (cyan), $eV_G/E_C=0.4$ (green), $eV_G/E_C=0.6$ (magenta) and $eV_G/E_C=0.8$ (dark green). The dashed red line corresponds to the geometric capacitance without the quantum correction. The first order approximation is no longer valid when $R_T/R_Q$ approaches unity. Inset shows the gate dependence of the total capacitance for $R_T/R_Q=2$.}
 \label{fig:capacitance}
\end{figure}

The current operator that we have used in the calculations takes only into account the tunneling of individual electrons. In problems involving a time-dependent driving it is also crucial to take into account the displacement current originating from the continuous displacement of electronic charge to satisfy the current conservation \cite{buttiker+pretre,bruder+schoeller}. We consider the following extension to our model: In addition to a ``particle current channel'' formed from the tunnel junctions we have a parallel ``displacement current channel'' formed from the capacitances of the tunnel junctions. We assume a left-right symmetric SET geometry with $C_G\ll C_L,\:C_R$. Because the gate capacitance is negligible compared to the other two capacitances, the current in the left lead equals the current in the right lead \cite{bruder+schoeller}. Thus the total admittance is given by a sum of the geometric and the tunneling contribution $G=G_Q-i\omega C/4$, where $G_Q$ is the admittance calculated from the Kubo formula. The quantum correction is usually smaller than the geometric contribution, but its gate dependence can still be observed provided that the tunneling resistance is not too large. The total capacitance of the SET, $C_\mathrm{tot}=C/4+\tilde{C}$, as a function of $R_T/R_Q$ is shown in Fig.~\ref{fig:capacitance}.

\begin{figure}
 \centering
 \includegraphics[width=6cm]{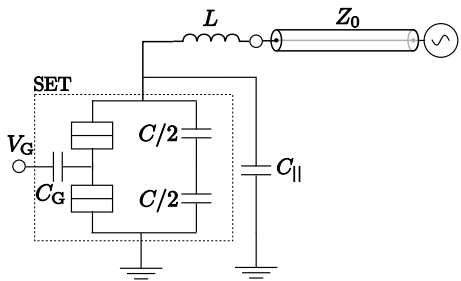}
 \caption{Proposed measurement setup for the measurement of gate charge with the reactive impedance of a SET.}
 \label{fig:schema}
\end{figure}

\begin{figure}
 \centering
 \includegraphics[width=8.5cm]{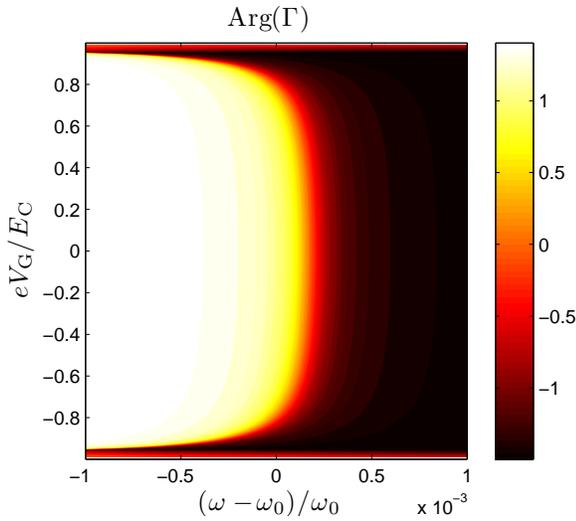}
 \caption{(color online) Phase of the reflection coefficient in a setup shown in Fig.~\ref{fig:schema}. Phase changes sign at the resonance frequency, which depends on the gate voltage. The parameters that were used are $R_T/R_Q=2$, $E_CL/\hbar R_Q=0.6$, $C_{||}/C=250$ and $Z_0/R_Q=0.002$. For $E_C/k_B=1\:\mathrm{K}$, these correspond to  $Z_0\approx50\:\Omega$, $L\approx160\:\mathrm{nH}$ and $C_{||}\approx250\:\mathrm{fF}$. Same values were used in the measurement of Ref.~\cite{sillanpaa}. The resonance frequency at zero gate is around $\omega_0\approx6\times10^9\:\mathrm{s}^{-1}$.}
 \label{fig:reflection}
\end{figure}

The reactive impedance can be utilized as an electrometer in a following setup (shown in Fig.~\ref{fig:schema}), similar to the rf-SET scheme: A resonator circuit formed of an inductor of inductance $L$, the SET and a stray capacitance $C_{||}$ is fed an rf-signal through a transmission line of impedance $Z_0$. The gate charge can then be probed by measuring the phase of the reflected signal. The phase of the reflection coefficient $\Gamma=(Z-Z_0)/(Z+Z_0)$, where $Z$ is the total impedance of the resonator circuit, as a function of frequency and gate voltage is shown in Fig.~\ref{fig:reflection}. Phase changes sign at the resonance frequency, which can be tuned by the gate voltage. Note that in typical rf-SET measurements so far the parameter range has been significantly different from those where this effect is observed, and thus it has stayed undetected.

Our analysis is limited by three factors. First of all, the exact charge state of the island is not calculated self-consistently, rather, it is assumed that the island is at all times in the ground state. This assumption should be valid at least when $|\hbar\omega|<\delta E^+_N,\delta E^-_N$, which is satisfied for most practical applications. Second, non-linear response effects are neglected within the applied linear response theory. Third, near charge degeneracy points and for $R_T/R_Q\lesssim 1$ second and higher order effects become relevant. However, none of the above mentioned limitations prevent the possibility to experimentally verify our analysis with currently available technology.

In conclusion, we have calculated a frequency-dependent SET admittance starting from the Kubo formula. We found novel features, most notably that the reactive part is nonzero under Coulomb blockade due to virtual processes and can be described by an effective capacitance at low frequencies. Our results can be experimentally demonstrated by rf-reflection measurements which should reveal a strong gate voltage and frequency dependence.

We thank D.~V. Averin, A.-P. Jauho, M.~A. Sillanp\"a\"a, P.~J. Hakonen and E.~B. Sonin for discussions. TTH is supported by the Academy of Finland.

\bibliography{setimpedance}

\end{document}